%% file: rpvlqd_arxiv_v2.tex
\documentclass[a4paper,11pt]{article}
\usepackage{jheppub}
\usepackage{breakurl}
\usepackage{mathrsfs}
\usepackage{amsmath}
\usepackage{amsfonts}
\usepackage{mathrsfs}
\usepackage{slashed}
\usepackage{epigraph}
\usepackage{fancyhdr}
\usepackage{amssymb}
\usepackage{graphicx}
\usepackage{longtable}
\usepackage{makecell}
\usepackage{amscd}
\usepackage{tabu}
\usepackage{bm}
\def\beq {\begin{equation}}
\def\eeq {\end{equation}}
\def\bea {\begin{eqnarray}}
\def\eea {\end{eqnarray}}

\def \MET{\rm E{\!\!\!/}_T}

\newcommand{\br}{\begin{eqnarray}}
\newcommand{\er}{\end{eqnarray}}
\newcommand{\be}{\begin{equation}}
\newcommand{\ee}{\end{equation}}

\usepackage{color}
\usepackage{hyperref} 
\hypersetup{linktocpage=true}
\usepackage[all]{hypcap}
\hypersetup{
   bookmarks=true,         
   unicode=false,          
   pdftoolbar=true,        
   pdfmenubar=true,        
   pdffitwindow=false,     
   pdfstartview={FitH},    
   pdftitle={My title},    
   pdfauthor={Author},     
   pdfsubject={Subject},   
   pdfcreator={Creator},   
   pdfproducer={Producer}, 
   pdfkeywords={keyword1} {key2} {key3}, 
   pdfnewwindow=true,      
   colorlinks=true,        
   linkcolor=blue,         
   citecolor=magenta,      
   filecolor=magenta,      
   urlcolor=cyan           
}
\usepackage{fancyhdr}

\begin{document}
\thispagestyle{empty}
\vspace*{-22mm}
\vspace*{10mm}
\hypersetup{backref=true,bookmarks}

\vspace*{10mm}

\begin{center}
{\Large {\bf\boldmath 
{Multi-leptons and Top-jets in the Hunt for Gluinos in R-parity Violating 
Supersymmetry}
}}

\vspace*{10mm}
{\bf Sanjoy Biswas$\rm ^{a,1}$, Diptimoy Ghosh$\rm ^{b,2}$, 
Saurabh Niyogi$\rm ^{c,3}$}\\
\vspace{4mm}
{\small
$\rm ^a$ Dipartimento di Fisica, Universit\'{a} di Roma ``La Sapienza'' \\
Piazzale A.  Moro 2, I-00185 Roma, Italy \\ [2mm]
$\rm ^b$ INFN, Sezione di Roma,\\
Piazzale A.  Moro 2, I-00185 Roma, Italy \\ [2mm]
$\rm ^c$ Regional Centre for Accelerator-based Particle Physics,\\
Harish-Chandra Research Institute,\\
Chhatnag Road, Jhusi, Allahabad - 211 019, India\\[2mm]
}
\vspace*{20mm}
{\bf Abstract}\vspace*{1mm}\\
\end{center}

The presence of R-parity ($\mathcal{R}_p$) violation offers intersting decay 
channels for the gluinos. In this work we present a new search strategy for 
the gluinos in the presence of semileptonic $\mathcal{R}_p$ violating couplings 
$\lambda^{'}_{133}$ and $\lambda^{'}_{233}$. We consider two scenarios 
(i) $\lambda^{'}$ induced 3-body decay of gluinos to a top quark ($t$), 
a bottom quark ($b$) and a light lepton ($\ell$) (ii) cascade decay of gluinos 
to top quarks and neutralinos ($\widetilde{\chi}_1^0$) followed by the decay of 
$\widetilde{\chi}_1^0$ to $t$, $b$ and $\ell$ through $\lambda^{'}$ couplings. 
We present two different search procedures which are common to both the  
scenarios. While the first one involves the traditional approach with 
multi-leptons and $b$-tagged jets, the second one employs the more recent 
technique to reconstruct highly energetic hadronically decaying top quarks. 
We perform a detailed simulation of the signal as well as all the relevant 
Standard Model backgrounds to show that the second procedure offers slightly 
better sensitivity for gluino discovery. In both the procedures, a $\geq$ 5$\sigma$ 
discovery is possible for the gluino mass in the range 1.5 -1.7 TeV 
at 14 TeV LHC with 50 fb$^{-1}$ integrated luminosity.

\vskip 130pt
\hrule
\small \noindent
\centerline{
$^1$sanjoy.biswas@roma1.infn.it \qquad
$^2$diptimoy.ghosh@roma1.infn.it \qquad
$^3$sourabh@hri.res.in} 
\newpage
\tableofcontents

\section{Introduction}

The discovery of the Standard Model (SM) higgs-like particle  
\cite{Aad:2012tfa,Chatrchyan:2012ufa} in the successful 8-TeV run of the Large 
Hadron Collider (LHC) has left us in a 
situation, as never before, where any significant excess of signal would now 
definitely point towards new physics (NP) beyond the SM. The Minimal 
Supersymmetric Standard Model (MSSM) 
\cite{Lykken:1996xt,Nilles:1983ge,Martin:1997ns,Haber:1984rc}, being one of 
the most attractive possibilities beyond the SM, would be the prime candidate 
to be searched for in various possible channels at the 14-TeV LHC. While 
Supersymmetry (SUSY) is clearly a broken symmetry, the demand of naturalness 
of the electroweak scale strongly suggests that at least some of the SUSY 
partners of the SM particles appear around a (few) TeV scale. The large amount 
of data collected at the 7 and 8-TeV run of the LHC, however, has already 
pushed the lower bounds of many of the SUSY partners well above a TeV. For 
example, the bound on the mass of the gluinos and the first two generation of 
squarks stand roughly around 1.5 TeV in the constrained version of the 
R-parity ($\mathcal{R}_p$) conserving MSSM \cite{Aad:2013wta,ATLAS-CONF-2013-061}. 
Indirect bounds from the measured value of the higgs boson mass and the low energy 
flavour observables also stand in the same ballpark 
\cite{Bhattacherjee:2010ju,Arbey:2011ab,Ghosh:2012dh,Dighe:2013wfa,Buchmueller:2013rsa}
\footnote{However, in general MSSM scenarios the bounds from higgs mass and flavour observables 
are much weaker.}.  

While in one hand this 
has made many theorists suspicious about the idea of naturalness, on the other 
hand this has also motivated many others to question the assumptions 
underlying the LHC searches, e.g., the assumption of $\mathcal{R}_p$ 
conservation. Note that the conservation of $\mathcal{R}_p$ renders the 
lightest SUSY particle (LSP) stable which often leads to large missing energy 
($\MET$) in the detectors. As large $\MET$ has been widely used by the 
experimental searches to suppress backgrounds, the $\mathcal{R}_p$ violating 
scenarios not only help evade the existing experimental constraints, but also 
open up a plethora of new possibilities and rich collider phenomenology.

Note that the lower bound on the mass of the top squark (stop) is rather model 
dependent even for the $\mathcal{R}_p$ conserving case; it's mass around a TeV 
(or a few hundred GeV less) is perfectly allowed in spite of the wealth of LHC 
data \cite{atlas_stop:2013,cms_stop:2013}. A large number of phenomenological 
studies has been devoted to stop searches both in the $\mathcal{R}_p$ 
conserving 
\cite{Ghosh:2012wb,Chakraborty:2013moa,Ghosh:2013qga,Belanger:2013oka, 
Drees:2013iya,Buckley:2012em,Carena:2013iba,Larsen:2012rq,Bi:2011ha} as well 
as $\mathcal{R}_p$ violating \cite{BM:RPV,Porod:2000pw,Han:2012cu, 
Evans:2012bf} scenarios. On the other hand, the bound on the gluino mass, in 
general, is stronger and the current lower value is more than a TeV even in 
most of the simplified scenarios (if the neutralino is not too heavy) 
\cite{Chatrchyan:2013lya,ATLAS-CONF-2013-061}. Hence, the $\mathcal{R}_p$ violating 
MSSM (thus, a very low MET) is 
an attractive possibility to realize TeV scale gluinos. This has motivated a 
number a studies focusing on signatures of gluino pair production in specific 
$\mathcal{R}_p$ violating scenarios 
\cite{Lisanti:2011tm,Hook:2012fd,Han:2012cu,Cohen:2012yc,
Berger:2013sir,Duggan:2013yna,Evans:2013jna}. 
For example, in reference \cite{Berger:2013sir} the authors 
considered the baryonic $\mathcal{R}_p$ violating couplings and obtained a 
lower bound of 800 GeV on the gluino mass using the CMS data in the same sign 
di-leption + $\MET$ + $b$-jet channel. These authors also reported an expected 
bound of 1.45 TeV on the gluino mass in their simplified scenario from the 14 
TeV LHC with 100 fb$^{-1}$ integrated luminosity. On the experimental front 
also many of the $\mathcal{R}_p$ violating scenarios are now being challenged; both 
the ATLAS and CMS collaborations have reported bounds on the masses of 
some MSSM particles in several $\mathcal{R}_p$ violating simplified 
scenarios \cite{Chatrchyan:2013xsw,CMS-PAS-SUS-13-010,ATLAS-CONF-2013-091,
Aad:2011kta,ATLAS-CONF-2012-153}.

In this work, we consider the presence of semileptonic $\mathcal{R}_p$ 
violating operators in the context of the gluino searches. We reiterate that 
the bound on the gluino mass is less model dependent (apart from the 
requirement of large MET) than the bound on stop mass as the production and 
decay of the gluinos mainly involve the QCD part of the MSSM lagrangian. 
However, in the decay cascade other parameters also come into play bringing in 
additional model dependence. In view of this, here we consider two simplified 
models consistent with the latest bounds. In the first case we assume that the 
top squark is heavier than the gluino ($m_{\widetilde{g}} < m_{\widetilde{t}_1}$)   
so that the $\widetilde{g}$ decays through $\tilde{g} \to t\, b\, \ell$ via an 
off-shell top squark ($\widetilde{t}_1$) (see section-\ref{model} and \ref{sig_bg} for 
details). In this case the top squark is required to be 
left-handed in order that it decays through the $\lambda^\prime$ coupling (see 
next section). 
Note that in this scenario the heavy-ness of the 
stop evades the possible bound from the $b$-$\ell$ resonance searches at the 
LHC \cite{Aad:2011ch,ATLAS:2012aq,Chatrchyan:2012vza,Chatrchyan:2012sv,ATLAS:2013oea}.
In the second scenario we assume the opposite hierarchy $m_{\widetilde{g}} > 
m_{\widetilde{t}_1}$ so that the gluinos decay though $\tilde{g} \to 
\widetilde{t}_1 (\to t\, \tilde{\chi}^0_1) \,t $. The $\tilde{\chi}^0_1$ now 
decays through $\tilde{\chi}^0_1 \to t \, b \, \ell$ in the presence of the 
same $\lambda^\prime$ coupling.

In both the cases the final state has top quarks, multiple leptons and 
$b$-jets. At first we investigate the traditional tri-leptons + 2$b$-jets 
final state. Note that we have chosen this final state in order that the same 
analysis can be applied to both the cases mentioned above. Proceeding further, 
we then point out that owing to the high mass of the gluinos some of the top 
quarks in the final state would often carry rather large transverse momentum 
and would give rise to collimated jets. We find that the use of 
jet-substructure techniques to tag these energetic ``top-jets'' are indeed 
very powerful to discover a signal.

The paper is organized as follows. In the next section we briefly discuss the 
simplified $\mathcal{R}_p$ violating scenario that we consider in our study. 
The details of signal and the backgrounds along with our search strategy is 
discussed in section-\ref{sig_bg}. In section-\ref{results} we present the 
summary of our simulation and discuss the results. Finally, we conclude in 
section-\ref{end}.

\section{RPV SUSY}
\label{model}

R-parity is a discrete symmetry of the MSSM lagrangian defined such a way that 
all the SM particles have {$\mathcal R$}-charge +1 and all their superpartners 
have {$\mathcal R$}-charge -1.  The above charge assignment allows 
$\mathcal{R}_p$ to be related to the Baryon ($B$) and Lepton ($L$) numbers by 
the following simple formula,

\begin{equation}
 \mathcal{R}_p = (-1)^{2S}  (-1)^{3(B-L)},
\end{equation}

where $S$ is the spin of the particle. The above formula makes it explicit 
that the conservation of $\mathcal{R}_p$ forbids all the dimension-4 and 5 
proton decay operators. As all the SUSY particles are odd under 
$\mathcal{R}_p$, this also makes the LSP stable providing a good dark matter 
candidate. 

However, it is possible that $\mathcal{R}_p$ is violated in specific ways that 
do not introduce fatal rates for proton decay. For example, switching on 
either $B$ or $L$ violating $\mathcal{R}_p$ violating couplings but not the 
both still forbids the dangerous proton decay operators.

In the absence of $\mathcal{R}_p$ the additional marginal and relevant terms 
allowed by gauge invariance in the MSSM superpotential can be written as 
\cite{Dreiner:1997uz,Schwanenberger:2004wv,Barbier:2004ez},

\begin{eqnarray}
W_{\mathcal{R}_p\hspace{-3mm}\slash} \supset \frac{1}{2} 
\lambda_{ijk} L_{i}L_{j}E^{c}_{k} + 
\lambda^{'}_{ijk} L_{i}Q_{j}D^{c}_{k} + 
\frac{1}{2} \lambda^{''}_{ijk} U^{c}_{i}D^{c}_{j}D^{c}_{k} + \mu_{i}L_{i}H_{u}
\label{rpv}
\end{eqnarray}

where $L_{i}$ ($E^{c}_{i}$) are left-handed lepton doublet (right-handed 
lepton) superfields, $Q_{i}$ ( $U^{c}_{i}$, $D^{c}_{i}$) the left-handed quark 
doublet superfields (right-handed Up-type and Down-type quark superfields, 
respectively) and $H_{u}$ is the Higgs superfield that gives mass to the 
up-type quarks.

Clearly, the couplings $\lambda^{''}$ violate $B$, while the couplings 
$\lambda$ and $\lambda^{'}$ violate $L$. As both $B$ and $L$ number must be 
violated to induce proton decay, it is still possible to turn on either 
$\lambda^{'}$ or $\lambda^{''}$ coupling without spoiling proton stability. 
Certain combinations of these $\mathcal{R}_p$ violating couplings are also 
constrained from various low energy observables (such as FCNC decays, 
neutron-antineutron oscillation, neutrino oscillation data, neutrino-less 
double beta decay, decays of tau lepton, meson mixing etc.) as well as data 
from high energy colliders e.g., LEP and Tevatron, see e.g., 
\cite{Bhattacharyya:1996nj} and the references therein.

In this work we consider the presence of $\lambda^{'}$ couplings in 
particular, $\lambda^{'}_{133}$ and $\lambda^{'}_{233}$, giving rise to the 
possibility of stop (coming from gluino decays in our case) decaying to a 
$b$-quark and light leptons (electron for $\lambda^{'}_{133}$ and muon 
for$\lambda^{'}_{233}$ ). Note that the $\lambda^{'}$-type couplings are 
also constrained to some extent by the various measurements mentioned above. 
However, the constraints on the couplings $\lambda^{'}_{133}$ and 
$\lambda^{'}_{233}$ are less severe as these involve the third generation 
quarks. The strongest bounds arise from the Majorana neutrino mass 
\cite{Bhattacharyya:1996nj,Godbole:1992fb} and flavor violating top decays 
\cite{Agashe:1995qm}. The $Z$-partial width also constrains these couplings
but to a lesser extent than the previous ones \cite{TAKEUCHI:2000dc}. 
However, these bounds are not strong enough to make the gluinos and 
neutralinos stable in the detectors.     

Although, ideally one should explain why some of these couplings are extremely 
small and some other are not, this is certainly quite interesting 
phenomenologically as it opens up many new decay channels of the MSSM 
particles leading to a rich phenomenology which should be studied in the 
colliders. As mentioned in the introduction, because of the absence of 
sufficient $\MET$ this also lowers the bounds on some of the SUSY particles. 

In the next section we will now discuss the specific decay topologies which are 
considered in this work along with the details of our simulation procedures.

\section{Signal, backgrounds and our search strategy} \label{sig_bg} 

\begin{figure}[h!]
\begin{center}
\includegraphics[scale=0.6]{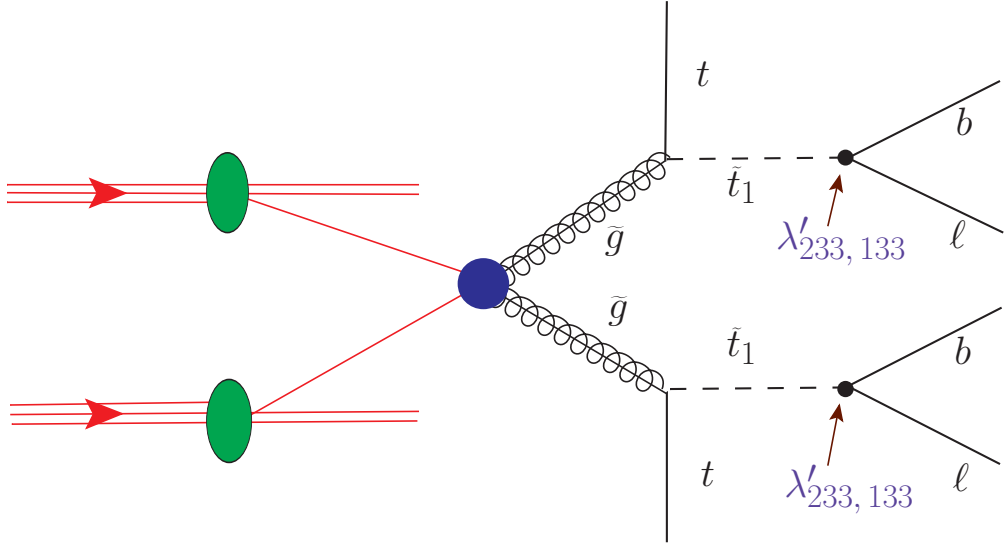}
\caption{The relevant Feynman diagram for gluino production followed by its decay 
to ($t\, b\, \ell$) final state in the presence of $\lambda^\prime$ couplings in 
our $\mathcal{R}_p$ violating scenario-1.
\label{fig:fmdiag1}}
\end{center}
\end{figure}

As mentioned in the introduction, in this work we consider gluino pair 
production followed by two different decay chains. In the first case, the 
$\widetilde{g}$ decays to ($t\, b\, \ell$) through an off-shell stop: 
$$
p  p  \to \widetilde{g} \; \widetilde{g}, \; \; \widetilde{g} \to t \; b \; \ell 
\; \; \; \; \textnormal{ (scenario - 1)}.
$$
A sample Feynman diagram is shown in Figure-\ref{fig:fmdiag1}. Looking at the 
structure of the $\lambda^\prime$ coupling (Eq.~\ref{rpv}) reveals that the top 
squark in this case has to have large left-handed component.

\begin{figure}[h!]
\begin{center}
\includegraphics[scale=0.5]{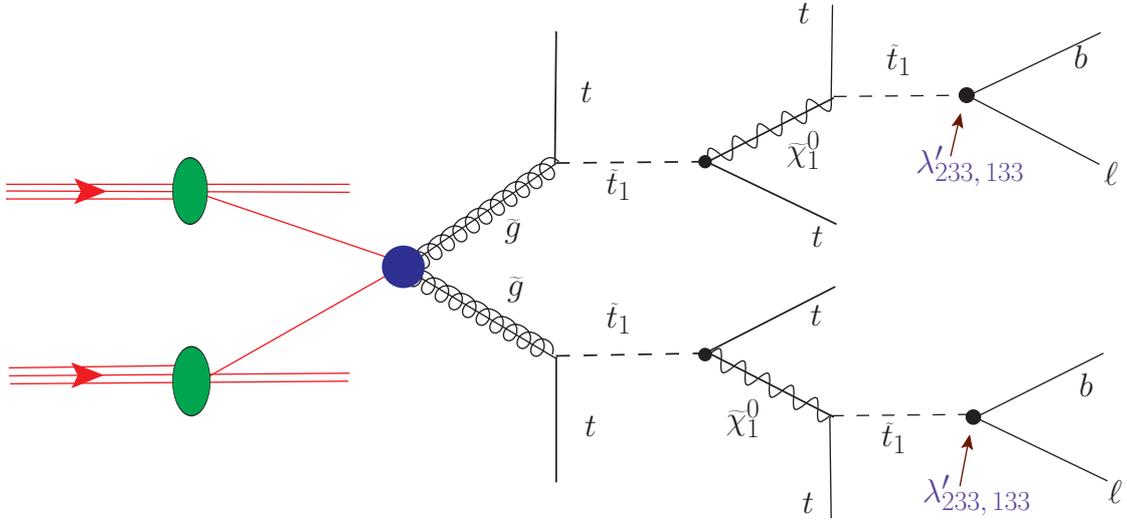}
\caption{The relevant Feynman diagram for gluino production followed by it's 
cascade decay to $t$, $b$ and leptons through $\lambda^\prime$ couplings in 
our $\mathcal{R}_p$ violating scenario-2.
\label{fig:fmdiag2}}
\end{center}
\end{figure}

In the second case we assume the decay chain,
$$
p \, p \to \widetilde{g} \, \widetilde{g}, \; \; 
\widetilde{g} \to \widetilde{t}_1 \, t,  \; \;
\widetilde{t}_1 \to t \, \widetilde{\chi}_1^0, \; \;
\widetilde{\chi}_1^0 \to t \, b \, \ell 
\; \; \; \; \textnormal{ (scenario - 2)}. 
$$
Figure-\ref{fig:fmdiag2} shows a sample Feynman diagram for this process where 
the $\widetilde{\chi}_1^0$ is assumed to decay through the left-handed component of 
an off-shell stop
\footnote{The same decay $\widetilde{\chi}_1^0 \to t \, b \, \ell$ 
can also proceed via an off-shell left-handed slepton. However, in that case the branching 
ratio can not be too high (typically less than 0.5) because of the presence of 
the left-handed sneutrino in the spectrum with the same mass (as the left-handed charged slepton), 
thus giving rise to also the decay  $\widetilde{\chi}_1^0 \to b \, \bar{b} \, \nu$. 
In fact, exactly these two decay chains (starting from stop pair production) 
were considered by the CMS collaboration to obtain a bound of about 700 
GeV on the stop mass if $\widetilde{\chi}_1^0$ is not too heavy \cite{Chatrchyan:2013xsw}.}.

In the following section we describe the details of our simulation procedure 
as well as the kinematic selection cuts for the signal and the relevant 
backgrounds. For simplicity, we will assume all the relevant branching ratios to 
be unity while presenting the results. For other values of the branching ratios 
our results can be easily scaled down appropriately.   

\subsection{Multilepton signal}
\label{multilep_cut}

We are now in a position to discuss the details of our event selection 
procedure. In this section we describe our analysis with multi-leptons and 
$b$-tagged jets. The analysis involving the tagging of top-jets will be 
discussed in the next subsection.

In the multi-lepton analysis, we first reconstruct the jets using the simple 
cone algorithm with the value of the radius parameter $R$=0.4. 
We consider only those jets which satisfy a transverse 
momentum cut $p_{T}^j \ge 20$ GeV and the pseudo-rapidity $|\eta| \le 
2.5$. Leptons are also selected with a transverse momentum $p_{T}^{\ell} 
\ge 10$ GeV and the pseudo-rapidity $|\eta| \le 2.5$. We call a lepton 
isolated if it satisfies \\

(i) The distance between the lepton and any of the jets $\Delta R (j \, \ell) 
> 0.4$\\

(ii) The distance between the lepton and any of the other leptons $\Delta 
R(\ell \, \ell) > 0.2$ \\

(iii)The ratio of the total hadronic transverse energy deposit within a cone of 
$\Delta R=0.2$ around the lepton to the lepton transverse energy is $\leq 
0.15$.
\newline
\newline
\indent
A jet is identified as a $b$-jet if it is close ($\Delta R < 0.2$) to a 
$b$-quark. For the $b$-tagging efficiency ($\epsilon_b$) we use the prescription 
from reference \cite{Chatrchyan:2012paa} which gives  $\epsilon_b$ = 0.71 
for $90 < p_T < 170$ GeV and 
at higher (lower) $p_T$ it decreases linearly with a slope of -0.0004 (-0.0047) 
GeV$^{-1}$. Moreover, the probability of 
mis-tagging a $c$-jet (light jet) as a $b$-jet is taken to be 20\% (0.73\%) 
\cite{ATLAS-CONF-2013-065}. 
Once the leptons and jets are constructed we use further analysis cuts for 
selecting events. Since the decay topologies in the two scenarios considered here 
are quite different, $p_T$ distributions of the observed leptons are expected to 
be different as well. In scenario-1, leptons are coming directly from the decay of a 
heavy gluino. Hence they are much more likely to pass harder $p_T$ cuts. 
On the other hand, leptons coming at a much later stage in the decay chain in 
scenario-2, are expected to be less energetic on the average. Therefore, it 
seems legitimate to employ different sets of lepton $p_{T}$ cuts for the two 
scenarios. We also reject (veto) an event if it has two opposite-charge-same-flavor 
leptons with invariant mass around the $Z$-boson mass.

We use the following set of selection criteria in scenario-1 :

\begin{itemize}
\item  Cut-I  : We demand that the event contains at least 3 isolated leptons. 
The three leptons must satisfy $p_{T_{\ell_1}} > $ 50 GeV, $p_{T_{\ell_2}} > $ 40 GeV 
and $p_{T_{\ell_3}} > $ 30 GeV. In addition, we also require that the event has at least 
2 jets.
\item Cut-II  : Veto on $Z$-boson. We choose the mass window to be $M_Z \pm 10$ GeV.
\item Cut-III  : The event must have at least two $b$-tagged jets.
\item Cut-IV : We define the effective mass of an event to be 
$M_{\rm eff} = \sum_{j} p_{T}^{j} + \sum_{\ell} p_{T}^{\ell} + \MET$ 
and demand that the event satisfies $M_{\rm eff} > $ 1000 GeV.
\end{itemize}

In scenario-2 we slightly change the $p_T$ cuts on the final state leptons. 
We demand the 3 leptons satisfy the $p_T$ cuts of 40, 30 and 20 GeV respectively 
instead of 50, 40 and 30 GeV as used in the previous case.

For the simulation of signal events we have used Pythia-6.4.24 
\cite{Sjostrand:2006za}. The Standard Model backgrounds have been generated 
using Alpgen v2.13 \cite{Mangano:2002ea} with the MLM prescription \cite{Hoche:2006ph} 
for the matching of matrix element hard partons and shower generated jets. We 
have used the CTEQ6L \cite{Pumplin:2002vw} parton distribution function for 
our simulations.

\subsection{Boosted top}
\label{top-jet}

In this subsection we describe our second search strategy which involves 
tagging an energetic top qurak using the jet substructure technique. We use the 
Johns Hopkins top tagger (JHTopTgger)\cite{Kaplan:2008ie} in this work. We now 
briefly discuss the steps of the JHTopTgger algorithm mentioning our choice of 
specific parameters as and when the occasion arises
\footnote{We have used the public package FastJet \cite{Cacciari:2005hq,Cacciari:2011ma} 
where this algorithm has been implemented.}. Here we closely follow 
the discussion in \cite{Kaplan:2008ie}.

\begin{figure}[h!]
\begin{center}
\begin{tabular}{cc}
\includegraphics[scale=0.75]{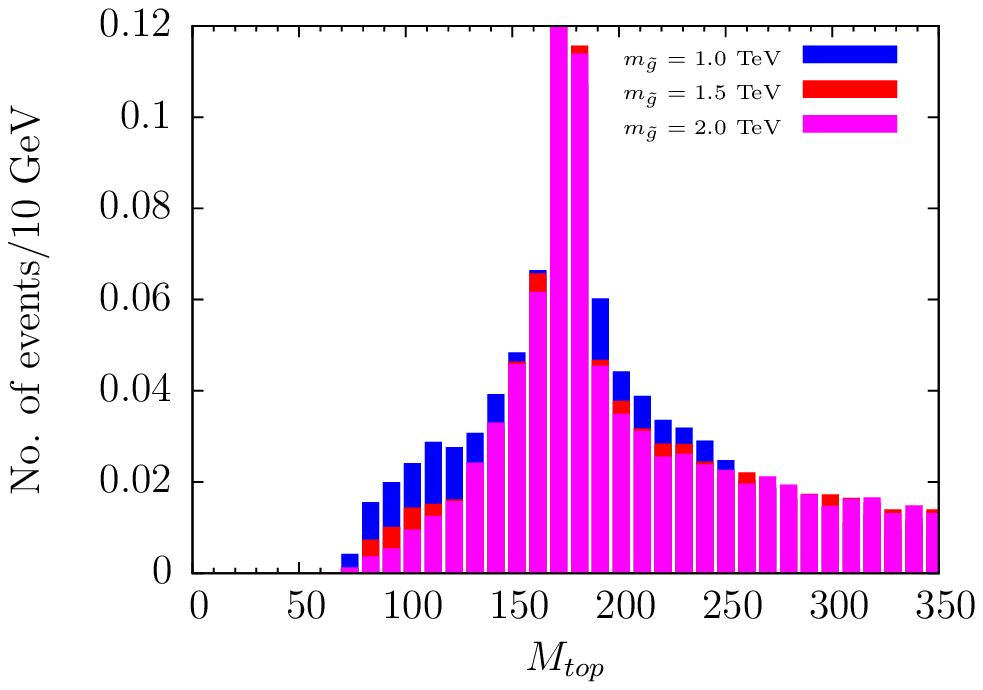}
\includegraphics[scale=0.75]{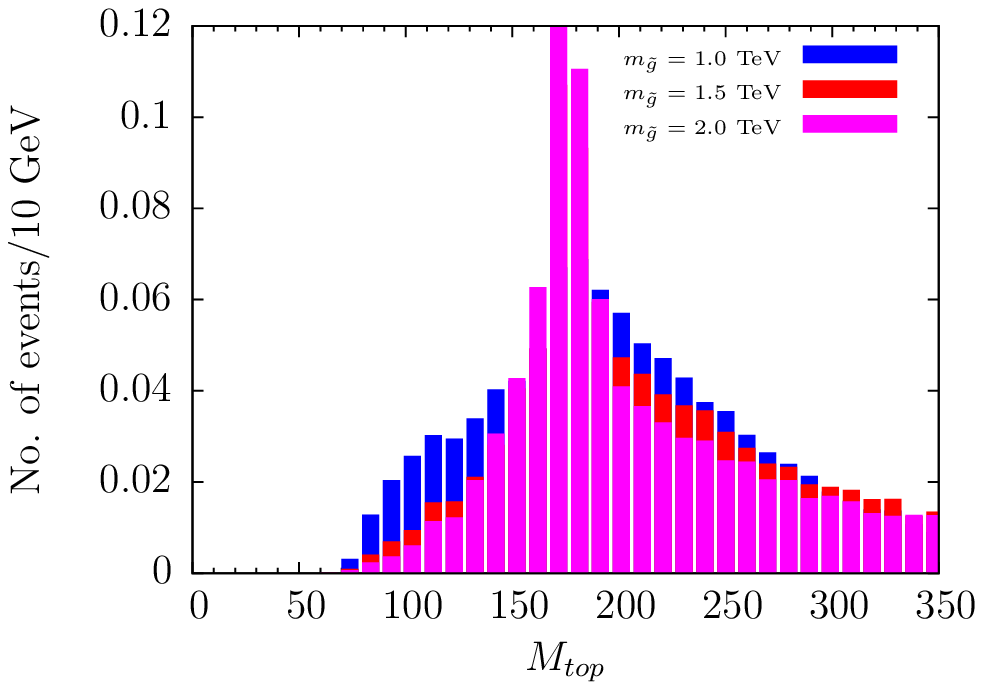}
\end{tabular}
\caption[]{The distribution of the mass of the reconstructed top quark 
from the hardest fat-jet in our $\mathcal{R}_p$ violating scenario-1 (left panel) 
and scenario-2(right panel) for the benchmark points chosen in Table-\ref{tab1}.
The total number of events has been normalized to unity.
\label{top-stop_reco}}
\end{center}
\end{figure}

\begin{enumerate} 

\item 
In the first step of the algorithm all the hadronic final states 
are clustered into the so-called `fat-jets' using the Cambridge-Aachen (CA) 
algorithm \cite{Dokshitzer:1997in} with the angular distance parameter 
$R = 1.0$. In the CA algorithm one starts with all the four-momenta of the 
hadronic final states and then combine the pair which has the smallest 
$\Delta R \equiv \sqrt{\Delta\eta^2 + \Delta\phi^2}$ (and $\Delta R < R$). 
The process is 
continued until there are no four-momenta left with $\Delta R < R$. The list 
of four-momenta which survive at this stage are then identified as 
jets.

\item 
In the second step of the algorithm one fat-jet (say, $J$) is considered 
at a time and it's 4-momenta $p^{(J)}$ is declustered into the 4-momenta 
($p^{(j1)}$ and $p^{(j2)}$) of the two subjets ($j1$ and $j2$) which were 
combined to get the fat-jet $J$. As the clustering history is stored at 
each stage of the CA algorithm, this step is same as reversing the clustering 
process mentioned in the previous paragaraph.

\item 
Three quantities are now computed, 
$$
\delta_1 = p_T^{(j1)}/p_T^{(J)}, \; \delta_2 = p_T^{(j2)}/p_T^{(J)} \; \textnormal{and} \;  
\delta = |\eta^{j1}-\eta^{j2}| +|\phi^{j1}-\phi^{j2}|. 
$$
The fat-jet $J$ is considered irreducible and having no substructure if the algorithm
encounters at least one of the following two situations,
\begin{itemize}
\item $\delta_1, \; \delta_2 < \delta_p$
\item $\delta < \delta_r$
\end{itemize}
Here  $\delta_p$ and $\delta_r$ are two adjustable parameters of the algorithm. 
In our analysis we set their values to $\delta_p = 0.1$ and $\delta_r = 0.2$.

\item 
If one of $\delta_1$ or $\delta_2$ comes out to be less than $\delta_p$, 
the corresponding subjet is discarded and the declustering procedure  
is applied to the other harder subjet. If both $\delta_1$ or $\delta_2$ are 
greater than $\delta_p$ then the declustering procedure is applied on both of them. 
This procedure stops when one of the conditions mentioned in step-3 is encountered 
or there is only one calorimeter cell left for the jet to be declustered. 

\item
The fat-jets with 3 or 4 subjets are kept for further analysis. The following three 
additional kinematic criteria are imposed before calling a fat-jet top tagged,
\begin{itemize}

\item 
The 3 or 4 subjets should reconstruct near the top quark mass. 
We use the top quark mass window to be $m_t \pm 20$ GeV. 

\item 
One pair of subjets (one of the combinations (1,2), (1,3), (1,4), (2,3), 
(2,4), (3,4)) should reconstruct near the $W$ mass. The $W$ mass window is 
chosen to be (60-100) GeV in our analysis.

\item 
The $W$ helicity angle $\theta_h$ satisfies cos$\theta_{h} < 0.7$.
The helicity angle is defined as the angle between the 3-momentum of the
reconstructed top quark and that of one of the the $W$ boson's decay 
products (the lower $p_T$ subjet in this case, as was proposed in the 
original work \cite{Kaplan:2008ie} also), 
as measured in the rest frame of the reconstructed $W$. 
 
\end{itemize}

Once the fat-jet satisfies the above kinematic cuts it is considered to be a 
true top candidate. Note that the subjet (or the hardest of the two subjets 
in case of four subjets) which is left after the $W$ reconstruction 
(we call this subjet as the non-$W$ subjet) is expected to be a $b$-jet 
for a true top candidate. In this work however, we do not demand that 
the non-$W$ subjet be $b$-tagged. In Figure-\ref{top-stop_reco} we 
show the reconstruction of top quark mass for our signal benchmark points 
(see Table-\ref{tab1}). 

\end{enumerate}

Once we find a top quark in an event (Cut-I) we then apply a few more selection criteria 
in order to combat the backgrounds. We discuss them below one by one. 

\begin{description}

\item[Cut-II] 
We demand at least two isolated leptons ($\ell1$ and $\ell2$) with $p_T > 50$ GeV. 
We use the same isolation criteria used in the previous section \ref{multilep_cut}.

\item[Cut-III] 
Apart from the $b$-jets coming from the top decays, the signal also has 
additional $b$-jets coming from gluino or neutralino decays. Keeping this mind we demand 
2 $b$-tagged jets ($b1$ and $b2$) which are far ($\Delta R > 0.8$) from 
the non-$W$ subjet of the reconstructed top quark. If an event has more than one 
reconstructed top quarks then only the hardest of them is used. 
The $b$-tagging procedure is again identical to that used in the previous subsection.

\item[Cut-IV] As a final selection criterion we demand that the effective mass of an event 
satisfies $M_{\rm eff} > 1250$ GeV. The definition of  $M_{\rm eff}$ is identical to that 
used in the previous section.

\end{description}

\section{Results and discussion}
\label{results}

%
\subsection{Multi-lepton signal}

In Table-\ref{tab1} and \ref{tab2}, we present the result of our analysis with 
$3\ell + 2b \, +$ jets (Section-\ref{multilep_cut}) for scenarios 1 and 2 respectively.
The first three columns show the processes studied, the raw production 
cross-section and the number of events generated for the signal and background 
processes. The raw production cross-sections for the signal points 
correspond to the next-to-leading-order value calculated using Prospino 
\cite{Beenakker:1996ed}
with default choices for the scale and the parton distribution function. For 
the background processes we use either the NLO cross-sections if they are 
available in the literature or the cross-sections obtained from Alpgen. 
For both the signal as well as the backgrounds the total number of events 
simulated is of the same order or more than the number expected in the 
14 TeV LHC with 50 fb$^{-1}$ luminosity (except for $t \, \bar{t}$ + jets). 
In the columns 4 - 6 the number of events after each selection cut 
(described in Section-\ref{multilep_cut}) are shown while the final column shows 
the cross-section after all the cuts have been imposed.

\begin{table}[!h]
\small
\begin{center}
\tabulinesep=1.2mm
\begin{tabu}{|l|cr|c|c|c|c|c|c|c|} 
\hline 
\hline
\multicolumn{4}{|}{} & 
\multicolumn{4}{|c|}{No. of events after the cut} & \\
\hline
Process & Production & & Simulated             
& C1 & C2 & C3 & C4 & Final cross- \\
        &  cross-section & & events &    
        &      &    &  & section (fb)   \\
\hline 
\multicolumn{9}{|c|}{Signal: $\tilde{g} \to t \, b \, \ell$ (scenario-1)} \\
\hline
$m_{\tilde{g}}$ = 1.0 &  370  fb &\cite{Beenakker:1996ed}& $5 \times 10^4$ 
& 7704   &  7427  &  2463  &  2412  & 17.85    \\ 
\hline
$m_{\tilde{g}}$ = 1.5 &  19  fb &\cite{Beenakker:1996ed}& $5 \times 10^4$ 
& 6971   &  6846  &  2185  &  2184 & 0.83     \\ 
\hline
$m_{\tilde{g}}$ = 2.0 &  1.56  fb &\cite{Beenakker:1996ed}& $5 \times 10^4$ 
& 6218  &  6116 & 1846  & 1846 & 0.058   \\ 
\hline
\multicolumn{9}{|c|}{Backgrounds} \\
\hline
$t \, \bar{t} \, + $ jets           & 953.6 pb &\cite{Czakon:2013goa}       
& 11607567  &  4  &  4  &   $<1$   & $<1$  &  0.0001    \\ 
$t \, \bar{t} \, Z \, + $ jets      & 1.121 pb &\cite{Kardos:2011na}        
&  140734   & 2667  & 842   &  306  & 19 &  0.15    \\ 
$t \, \bar{t} \, W \, + $ jets  & 769 fb   &\cite{Campbell:2012dh}          
& 169973   &  447  &  394   &  226 &  9  &  0.04   \\ 
$Z \, Z \, W \, +$ jets      &  44.3 fb &  \cite{Mangano:2002ea}                                
&  87650   &  272  &   52   &   4  &  1   &  0.0005    \\ 
$W \, W \; Z \, + $ jets       &  137.5 fb  &\cite{Mangano:2002ea}                               
&  65090   &  411  &   116    &   1  &  $<1$   &  -    \\ 
$W \, W \; W \, + \geq $ 2\, jets       & 94.1 fb  &\cite{Mangano:2002ea}                              
&  26268   &   18  &  8 &  $<1$  &  $<1$   &   -   \\ 
\hline 
Total & \multicolumn{7}{|c|}{} &  \\
Background & \multicolumn{7}{|c|}{} & 0.190 \\
\hline 
\hline
\end{tabu}
\caption{Event summary after individual selection cuts both for the MSSM benchmark 
points (in scenario-1) as well as the SM backgrounds for the multi-lepton analysis. 
The final cross-sections 
after all the selection cuts are shown in the last column. All the masses are 
in TeV.}
\label{tab1} 
\end{center}
\end{table} 

 \begin{figure}[h!]
 \includegraphics[width=210pt,height=180pt]{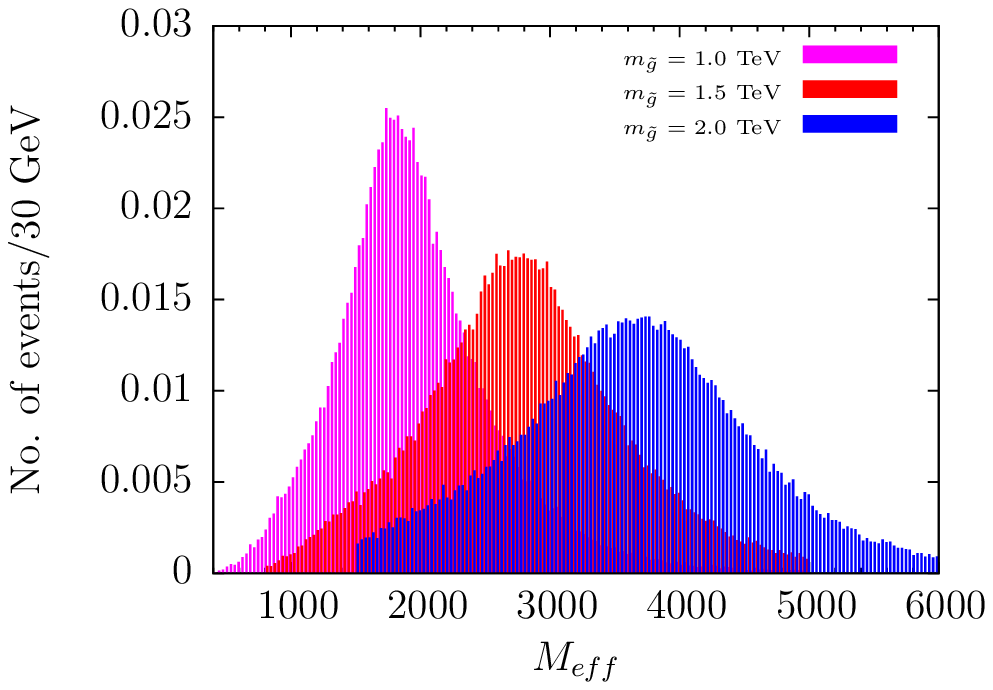}
 \includegraphics[width=210pt,height=180pt]{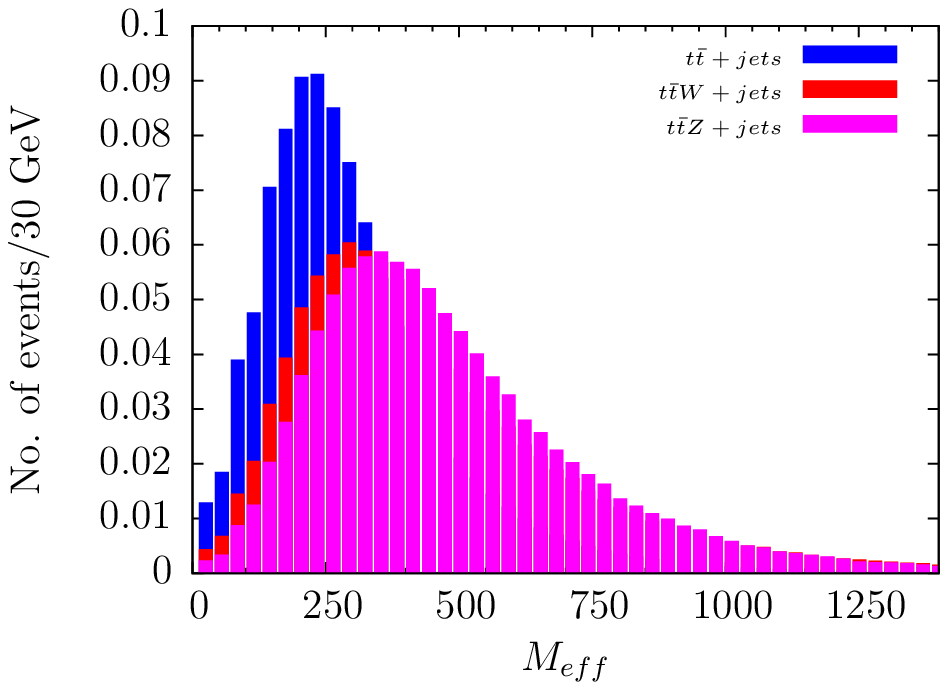}
 \caption{$M_{\rm eff}$ distribution for three signal benchmark points as well as 
 the dominant backgrounds. The total number of events has been normalized to unity.} 
 \label{fig:meff1}
 \end{figure}

The dominant backgrounds in this case are $t\bar{t}Z+$jets, $t\bar{t}W+$jets and 
$t\bar{t}+$jets. 
In case of $t\bar{t}+$jets the semileptonic decay of the $B$-mesons 
(in addition to the two leptons from the $W$'s) can contribute to the tri-lepton 
final state. 
Although the number of these events can be reduced to a good extent 
by requiring the isolation criteria as described in Section-\ref{multilep_cut}, the 
enormous cross-section of the $t\bar{t}+$jets process makes them significant background unless
we impose hard lepton $p_T$ cuts. In fact, we have been able to reduce events from 
$t\bar{t}+$jets drastically after imposing 50 GeV, 40 GeV and 30 GeV cuts respectively on first 
three hardest leptons (see Table-\ref{tab1}). Some events remain in scenario-2 
(see Table-\ref{tab2}) where we lessen the $p_T$ cuts on leptons.
The other dominant (the leading one in scenario-1) contribution comes from 
the $t \, \bar{t} \, Z \, +$ jets background where additional 
leptons can originate from the $Z$ boson decay. A $Z$ veto reduces this 
background substantially. 
The process $t\bar{t}W+$jets also constitutes 
a background to our signal but its contribution is much less than 
$t \, \bar{t} \, Z \, +$ jets as can be seen from Table-\ref{tab1} and \ref{tab2}. 
We have also generated other potential backgrounds e.g., 
$Z \, Z \, W \, +$ jets, $W \, W \; Z \, +$ jets and 
$W \, W \; W \, +$ jets whose contributions 
are negligible compared to the previous ones when a $Z$ veto and the requirement of 
2 $b$-jets are also imposed. The use of the effective mass in the final step of the algorithm reduces 
backgrounds quite efficiently without affecting the signal almost at all. As the 
effective mass of an event for the signal is closely related to $2 m_{\widetilde{g}}$ 
the distribution peaks at very high values of the effective mass, see Figure-
\ref{fig:meff1}a where we show the effective mass distributions for three of our 
benchmark points with different choices of the gluino mass (in scenario-1). 
In Figure-\ref{fig:meff1}b we present the same distribution for the dominant backgrounds.

\begin{table}[!h]
\small
\begin{center}
\tabulinesep=1.2mm
\begin{tabu}{|l|cr|c|c|c|c|c|c|c|} 
\hline 
\hline
\multicolumn{4}{|}{} & 
\multicolumn{4}{|c|}{No. of events after the cut} & \\
\hline
Process & Production & & Simulated             
& C1 & C2 & C3 & C4 & Final cross- \\
        &  cross-section & & events &    
        &      &    &  & section (fb)   \\
\hline 
\multicolumn{9}{|c|}{Signal: $\widetilde{g} \to \widetilde{t}_1 \, t, \; 
\widetilde{t}_1 \to t \, \widetilde{\chi}_1^0, \; \widetilde{\chi}_1^0 \to t \, b \, \ell$ 
(scenario-2)} \\
\hline
$(m_{\tilde{g}}, m_{\tilde{t}})$ = (1.0, 0.8) &  370  fb &\cite{Beenakker:1996ed}& $5 \times 10^4$ 
& 10609   &  8718  &  5663  &  5491  & 40.63    \\ 
\hline
$(m_{\tilde{g}}, m_{\tilde{t}})$ = (1.5, 1.0) &  19  fb &\cite{Beenakker:1996ed}& $5 \times 10^4$ 
& 10241   &  8794  &  6060  &  6058 & 2.3     \\ 
\hline
$(m_{\tilde{g}}, m_{\tilde{t}})$ = (2.0, 1.0) &  1.56  fb &\cite{Beenakker:1996ed}& $5 \times 10^4$ 
& 8706  &  7554 &  4823  & 4823 & 0.15   \\ 
\hline
\multicolumn{9}{|c|}{Backgrounds} \\
\hline
$t \, \bar{t} \, + $ jets           & 953.6 pb &\cite{Czakon:2013goa}       
& 11607567  &  316  &  300  &   87   & 10  &  0.82    \\ 
$t \, \bar{t} \, Z \, + $ jets      & 1.121 pb &\cite{Kardos:2011na}        
&  140734   & 2870  & 887   &  350  & 21 &  0.167    \\ 
$t \, \bar{t} \, W \, + $ jets  & 769 fb   &\cite{Campbell:2012dh}          
& 169973   &  594  &  512   &  272 &  15  &  0.067   \\ 
$Z \, Z \, W \, +$ jets      &  44.3 fb &  \cite{Mangano:2002ea}                                
&  87650   &  320  &   34    &   3  &  1   &  0.0005    \\ 
$W \, W \; Z \, + $ jets       &  137.5 fb  &\cite{Mangano:2002ea}                               
&  65090   &  467  &   162    &   2  &  1   &  0.002    \\ 
$W \, W \; W \, + \geq $ 2\, jets       & 94.1 fb  &\cite{Mangano:2002ea}                              
&  26268   &   34  &  23 &  $<1$  &  $<1$   &   -   \\ 
\hline 
Total & \multicolumn{7}{|c|}{} &  \\
Background & \multicolumn{7}{|c|}{} & 1.05 \\
\hline 
\hline
\end{tabu}
\caption{Event summary after individual selection cuts both for the 
MSSM benchmark points (in scenario-2) as well as the SM backgrounds for the multi-lepton analysis. 
The final cross-sections 
after all the selection cuts are shown in the last column. All the masses are 
in TeV. We have taken the mass of $\widetilde{\chi}_1^0$ to be 300 GeV in this case.}
\label{tab2} 
\end{center}
\end{table} 

%
\subsection{Boosted top}

In this subsection we present our results using the techniques described in 
Section-\ref{top-jet}. The summary of our findings is shown in Table-\ref{tab3} where 
the same conventions as in Table-\ref{tab1} and \ref{tab2} have been used. 
In columns 4 - 7 the number of events after each selection cut 
(see Section-\ref{top-jet}) are shown. The final column shows the final cross-section 
after all the selection cuts.

\begin{table}[h!]
\small
\begin{center}
\tabulinesep=1.2mm
\begin{tabu}{|l|cr|c|c|c|c|c|c|c|} 
\hline 
\hline
\multicolumn{4}{|}{} & 
\multicolumn{4}{|c|}{No. of events after the cut} & \\
\hline
Process & Production & & Simulated             
& C1 & C2   & C3 & C4   & Final cross-  \\
        &  cross-section & & Events &    
        &      &    &    &    section (fb)   \\
\hline 
\multicolumn{9}{|c|}{Signal: $\tilde{g} \to t \, b \, \ell$ (scenario-1) } \\
\hline
$m_{\tilde{g}}=$ 1.0 &  370  fb &\cite{Beenakker:1996ed}& $5 \times 10^4$
&   7855  &  5257  & 2690   &  2681   & 19.83   \\ 
\hline
$m_{\tilde{g}}=$ 1.5 &  19  fb &\cite{Beenakker:1996ed}& $5 \times 10^4$
&  9007  &  6290  &  3044   &  3044   & 1.16  \\ 
\hline
$m_{\tilde{g}}=$ 2.0  &  1.56 fb &\cite{Beenakker:1996ed}& $5 \times 10^4$ 
&  9297    &  6521   &  2958   &  2958  &  0.09  \\ 
\hline
\multicolumn{9}{|c|}{Signal: $\widetilde{g} \to \widetilde{t}_1 \, t, \; 
\widetilde{t}_1 \to t \, \widetilde{\chi}_1^0, \; \widetilde{\chi}_1^0 \to t \, b \, \ell$ 
(scenario-2)} \\
\hline
$(m_{\tilde{g}}, m_{\tilde{t}})$= (1.0, 0.8) &  370  fb &\cite{Beenakker:1996ed}& $5 \times 10^4$ 
& 11982  & 3819  & 3023  &  2842  & 21.03  \\ 
\hline
$(m_{\tilde{g}}, m_{\tilde{t}})$= (1.5, 1.0) &  19  fb &\cite{Beenakker:1996ed}& $5 \times 10^4$ 
& 15935   & 5438   &  4545  &  4534  &  1.72  \\ 
\hline
$(m_{\tilde{g}}, m_{\tilde{t}})$= (2.0, 1.0) &  1.56  fb &\cite{Beenakker:1996ed}& $5 \times 10^4$ 
&  18130    & 5954   &  4973  & 4972  &   0.155  \\ 
\hline
\multicolumn{9}{|c|}{Backgrounds} \\
\hline
$t \, \bar{t} \, + $ jets           & 953.6 pb  &\cite{Czakon:2013goa}   &
31712564 & 424944 & 166 & 20 & 4 &  0.12 \\ 
$t \, \bar{t} \, Z \, + $ jets      & 1.121 pb  &\cite{Kardos:2011na}    &
226110 & 9105 & 210 & 12 &  1 &  0.005  \\ 
$t \, \bar{t} \, W \, + $ jets  & 769 fb    &\cite{Campbell:2012dh}  &
276807 & 12105 & 164 & 10 &  1  &  0.003  \\ 
$t \, \bar{t} \, h \, + $ jets      & 700 fb    &\cite{Dawson:2003zu}    &
231064 & 10926 & 67 & 8 &  1  &  0.003  \\ 
\hline 
Total & \multicolumn{7}{|c|}{} &  \\
Background & \multicolumn{7}{|c|}{} & 0.131\\
\hline 
\hline
\end{tabu}
\caption{Event summary after individual selection cuts both for 
the MSSM benchmark points as well as the SM backgrounds for the boosted top analysis. 
The final cross-sections after all the selection cuts are shown 
in the final column. All the masses are in TeV.  We have taken the 
mass of $\widetilde{\chi}_1^0$ to be 300 GeV in scenario-2.
\label{tab3}} 
\end{center}
\end{table} 

A comparison of columns 3 and 4, both for the signal and the backgrounds, 
clearly reveals the effectiveness of tagging an energetic top quark. In case of 
the $t \, \bar{t}$ +jets background the number of events get reduced by almost a factor 
of 75 while for the signal the loss is only by a factor of 5. Such a large 
gain is due to the large average transverse energy (hence collimated jets) of 
the top quark in the signal compared to the background. For the background 
events the jets from the decay of a top quark are highly separated from each other 
and are not captured by a fat-jet which in turn reduces the tagging efficiency. The demand 
of two isolated leptons as well as two $b$-tagged jets also helps tame the 
background to a large extent. Note that these $b$-jets are far from the non-$W$ 
candidate of the tagged top quark (see Section-\ref{top-jet} for details). This criteria 
makes sure that the $b$-jets, most of the time, are not from the decay of 
the top quark which has been reconstructed.

In the last step of the analysis, the $M_{\rm eff}$ cut brings down the background 
to a minuscule level keeping a handful of signal events. We have checked that the 
processes $t \, Z$ + jets,  $t \, W$ + jets and $t \, h$ + jets do not contribute 
to our final number of background events.

\subsection{Comparison of the two analyses}

In Table-\ref{tab:sig} we compute the signal significance obtained from the two 
analyses described above and compare them. We define the significance $\sigma$ 
as $\sigma = S/\sqrt{S+B}$ where $S$ and $B$ are the absolute number of signal 
and background events respectively for a particular luminosity.
At first we should mention that the raw signal cross-section reduces very fast 
as the gluino mass is increased; the cross-section drops to 1.56 fb for a gluino mass 
of 2 TeV from a healthy 370 fb for a gluino mass of 1 TeV. This is the primary 
reason for the significance to drop dramatically with increasing $\widetilde{g}$ mass, as 
can be seen from Table-\ref{tab:sig}. 
\begin{table}[h!]
\begin{center}
\tabulinesep=1.2mm
\begin{tabu}{|c||c|c|c|c|c|c|} 
\hline
\hline
Gluino mass & \multicolumn{2}{|c|}{Scenario-1} &  
               \multicolumn{2}{|c|}{Scenario-2}  \\ 
\cline{2-5}
 ~(in GeV)     & multi-lepton   & boosted-top  &   multi-lepton  & boosted-top   \\
\hline
 1000       & 29.71 (42.02)    & 31.38 (44.38) &  44.5 (62.9) &  32.32 (45.71)    \\
\hline
1500        &  5.81 (8.22)     &  7.21 (10.20) &  8.88  (12.56)  &  8.93 (12.64)    \\
\hline
2000        &  0.82 (1.16)     & 1.35 (1.91)  &  0.96  (1.37)   &  2.04 (2.89)  \\
\hline
\hline
\end{tabu}
\caption{The statistical significance of our signal for the two analysis strategies 
discussed in the text. The numbers outside (inside) the parentheses corresponds to 
an integrated luminosity of 50 fb$^{-1}$ (100 fb$^{-1}$).
\label{tab:sig}} 
\end{center}
\end{table}

We present the significance assuming two integrated luminosities, 50 fb$^{-1}$ 
and 100 fb$^{-1}$ (the numbers within the parentheses). It can be seen that 
for our first analysis strategy the significance is more than 8 (12) for the 
$\mathcal{R}_p$ violating scenario-1 (scenario-2) with $\widetilde{g}$ 
mass of 1.5 TeV and assuming 100 fb$^{-1}$ data set. Even at 50 fb$^{-1}$ integrated 
luminosity the significance $ \sim 6 (9)$ is achieved in scenario-1 (scenario-2). 
For a 2 TeV $\widetilde{g}$ mass the significance drops dramatically as can be seen from the 
last row of Table-\ref{tab:sig}.

On the other hand, for the second analysis strategy the significance is about 
7.2 (compared to about 5.8 in the multilepton case ) in the scenario-1 for the 
$\widetilde{g}$ mass of 1.5 TeV and 50 fb$^{-1}$ integrated luminosity. In scenario-2, 
however, both the analysis give similar results. 

It is worth mentioning here that while calculating the significance we have not taken into 
account any systematic uncertainty which is very difficult to estimate in a reliable way. 
We believe that for the cases where the significance is large enough (say, $\geq 5$) 
the effect of including systematic uncertainties should not be large. For example, 
in the multilepton analysis in scenario-1 with gluino mass 1.5 TeV, adding a 30\% systematic 
uncertainty on the background crosssection reduces the significance from 5.81 (8.22) 
to 5.65 (7.99) for 50 fb$^{-1}$ (100 fb$^{-1}$) integrated luminosity.

\section{Conclusion}
\label{end}

Introduction of $\mathcal{R}_p$ violation in the MSSM lagrangian is a phenomenologically 
attractive way to evade strong bounds on the masses of SUSY particles obtained in 
the R-parity conserving scenario. This has fueled a significant amount of effort from 
both the experimentalists and theorists in investigating new signal topologies 
present in the R-parity violating case. However, there exists many more potentially 
interesting possibilities which still need to be covered. In this work, we have 
concentrated on two such topologies which exist in the presence of semileptonic 
R-parity violation with $\lambda^\prime_{133, \, 233}$ couplings. We have considered 
gluino pair production with their subsequent decay to top quarks, leptons and $b$-jets.   
Two analysis strategies have been considered, one with the canonical multi-leptons 
and $b$-jets and the other one with more recent technique to reconstruct highly 
energetic top quarks. We performed a detailed simulation of the signal and all possible  
background processes to estimate and compare the effectiveness of these two procedures.

In order to present our results in a clear way we have chosen a few benchmark 
scenarios for both the topologies considered. Our results have been summarized in
Table-\ref{tab1}, \ref{tab2} and \ref{tab3} and a comparison of the effectiveness of the two 
analyses procedures have been presented in Table-\ref{tab:sig}. 
We observe that our second strategy which involves reconstructing a top
quark from the final state was slightly more effective compared to the traditional 
multi-leptons + $b$-jets analysis. While in the multi-lepton with $b$-jets search
the significance can reach up to as high as 5.8 (8.9) for a gluino mass of 1.5 TeV in 
the $\mathcal{R}_p$ violating scenario-1 (scenario-2) with a 50 fb$^{-1}$ data set, the 
second analysis does somewhat better in scenario-1 (significance rises to 7.2) and provides 
comparable sensitivity ($\sim 9$) in the second scenario.

Note that, although we present our results in two simplified scenarios just 
for clear illustration of the procedures, our analysis can be applied to any other 
situations with similar final state. We would also like to mention that we have 
not considered any detector effects and pile-up contaminations which are rather difficult 
for us to simulate in a reliable way. However, we believe that our analysis can be taken 
as a guiding reference for more detailed and realistic analysis on real data 
by our experimental colleagues.

\section*{Acknowledgements}

The research leading to these results has received funding from the European 
Research Council under the European Union's Seventh Framework Programme 
(FP/2007-2013) / ERC Grant Agreement n.279972. SB and DG would like to thank 
the members of the High Energy Theory Group at the Sapienza University of Rome 
for useful discussions. SN likes to acknowledge the 
computing facility at the Regional Centre for Accelerator-based Particle 
Physics (RECAPP), Harish-Chandra Research Institute. SN is also thankful to 
University of Helsinki and Helsinki Institute of Physics for the hospitality 
where the final part of the project has been carried out.

\input{rpvlqd_arxiv_v2.bbl}
\end{document}

%% file: rpvlqd_arxiv_v2.bbl
\providecommand{\href}[2]{#2}\begingroup\raggedright\endgroup